\documentclass[pra,twocolumn,showpacs,superscriptaddress]{revtex4}

\usepackage{graphicx}
\usepackage{amsmath}
\usepackage{amssymb}

\newcommand{\beq}{\begin{equation}}
\newcommand{\eeq}{\end{equation}}
\newcommand{\beqa}{\begin{eqnarray}}
\newcommand{\eeqa}{\end{eqnarray}}
\newcommand{\beqan}{\begin{eqnarray*}}
\newcommand{\eeqan}{\end{eqnarray*}}

\begin{document}
\title{Effective squeezing enhancement 
via measurement-induced non-Gaussian operation 
and its application to dense coding scheme}
\author{Akira Kitagawa}
\affiliation{National Institute of Information and Communications Technology (NICT) 
4-2-1 Nukui-Kita, Koganei, Tokyo 184-8795 Japan}
\affiliation{Core Research for Evolutional Science and Technology (CREST), 
Japan Science and Technology Agency \\
1-9-9 Yaesu, Chuoh, Tokyo 103-0028 Japan}
\author{Masahiro Takeoka}
\affiliation{National Institute of Information and Communications Technology (NICT) 
4-2-1 Nukui-Kita, Koganei, Tokyo 184-8795 Japan}
\affiliation{Core Research for Evolutional Science and Technology (CREST), 
Japan Science and Technology Agency \\
1-9-9 Yaesu, Chuoh, Tokyo 103-0028 Japan}
\author{Kentaro Wakui}
\affiliation{National Institute of Information and Communications Technology (NICT) 
4-2-1 Nukui-Kita, Koganei, Tokyo 184-8795 Japan}
\affiliation{Core Research for Evolutional Science and Technology (CREST), 
Japan Science and Technology Agency \\
1-9-9 Yaesu, Chuoh, Tokyo 103-0028 Japan}
\affiliation{Department of Applied Physics, The University of Tokyo \\
7-3-1 Hongo, Bunkyo-ku, Tokyo 113-8656 Japan}
\author{Masahide Sasaki}
\email{psasaki@nict.go.jp}
\affiliation{National Institute of Information and Communications Technology (NICT) 
4-2-1 Nukui-Kita, Koganei, Tokyo 184-8795 Japan}
\affiliation{Core Research for Evolutional Science and Technology (CREST), 
Japan Science and Technology Agency \\
1-9-9 Yaesu, Chuoh, Tokyo 103-0028 Japan}

\date{\today} 

\begin{abstract}
We study the measurement-induced non-Gaussian operation 
on the single- and two-mode \textit{Gaussian} squeezed vacuum states 
with beam splitters and on-off type photon detectors, 
with which \textit{mixed non-Gaussian} states are generally 
obtained in the conditional process. It is known that 
the entanglement can be enhanced via this non-Gaussian operation 
on the two-mode squeezed vacuum state. 
We show that, in the range of practical squeezing parameters, 
the conditional outputs are still close to Gaussian states, 
but their second order variances of quantum fluctuations 
and correlations are effectively 
suppressed and enhanced, respectively. 
To investigate an operational meaning of these states, 
especially entangled states, we also evaluate 
the quantum dense coding scheme from the viewpoint 
of the mutual information, and we show that 
non-Gaussian entangled state can be advantageous 
compared with the original two-mode squeezed state.

\end{abstract}
\pacs{03.67.Hk, 03.67.Mn, 42.50.Dv}

\maketitle 

\section{Introduction}

Non-classical optical Gaussian states, 
such as single- or two-mode squeezed vacuum states, 
play essential roles in continuous variable (CV) 
quantum information technology. 
These states have already been implemented in labs 
and manipulating 
these states with linear operations including beam splitters, phase 
shifters, displacements, and homodyne measurements, various 
quantum information protocols have been demonstrated: 
teleportation \cite{Furusawa98}, dense coding \cite{Mizuno05}, and 
entanglement swapping \cite{Jia04}. 
Mathematically, the operations by these linear optics tools and 
arbitrary squeezing operations are described by linear and bilinear 
Hamiltonians and classified as {\it Gaussian operations} 
which transform Gaussian states into Gaussian states.

A class of Gaussian operation is, however, obviously a part of 
the class of universal quantum operations. 
Recent theoretical investigation has shown the limitation of 
Gaussian operations in quantum information processing, 
e.g. entanglement distillation of Gaussian input to Gaussian output 
is impossible \cite{Eisert02,Fiurasek02,Giedke02}
and, more generally, quantum information protocols 
consisting of only Gaussian operations can be simulated classically
\cite{Bartlett02PRL}. 
Therefore, implementation of {\it non-Gaussian operation}, 
that is the operations accessible to the outside of the Gaussian domain, 
would be crucial to extract ultimate potential of quantum information theory. 
In addition, a delightful theoretical result is that, in principle, 
arbitrary operations can be implemented by combining one of non-Gaussian 
operations with suitable Gaussian operations \cite{Lloyd99,Bartlett02PRA}.

Although, at present, even the cubic nonlinearity is hard to realize 
on the level of single photon, 
there is an alternative idea, called the measurement-induced nonlinearity, 
where effective nonlinearity is associated with non-Gaussian measurement 
such as photon counting \cite{Bartlett02PRA}. 
A simplest example of such operations 
has been theoretically investigated, where photons 
in non-classical Gaussian states are subtracted 
by low reflectance beam splitters and photon counters. 
It was proposed that one can generate Schr\"{o}dinger cat-like state 
from a single-mode squeezed vacuum by subtracting photons \cite{Dakna97}. 
Recently, nonclassicality of this cat-like state 
has been investigated with respect to negativity of Wigner function, 
in consideration of some experimental parameters
\cite{Kim05}.

Moreover, it is predicted that the photon subtraction 
from two-mode squeezed vacuum can increase entanglement 
\cite{Opatrny00,Cochrane02,Browne03}. 
In ideal situation, i.e. perfect photon number counting and lossless setup, 
the output is always pure and one can uniquely quantify 
the increase of entanglement by the von Neumann entropy of a partial system. 
In practical situation, however, imperfections should be taken into account. 
The main one is the imperfection of photon detector. 
It is still difficult to distinguish the photon number precisely. 
Currently available type of detector is the on-off type detector 
(e.g. avalanche photodiodes in Geiger mode operation)
which discriminate only between the vacuum and the presence of photons 
with finite quantum efficiency and nonzero dark counts. 
This type of detector suffices for some purposes. 
In fact, the first observation of non-Gaussian statistics 
due to the photon subtraction from a single-mode squeezed vacuum 
was demonstrated using such a type of device \cite{Wenger04}.

A serious restriction due to the on-off type resolution is that 
the detector projects the original state into a mixed state. 
For instance, in the case where one of the two mode entangled state 
is measured by the on-off detector, the state of the other mode 
is projected into a mixed state. Similarly, 
the photon subtraction with on-off detector reduces 
the pure two-mode squeezed state to the mixed state. 
One cannot, therefore, apply the von Neumann entropy 
to quantify their entanglement. 
In this direction, some operational measures have been exploited 
theoretically to characterize the non-Gaussian mixed entangled state. 
They are, for example,  
the improvement of the teleportation fidelity \cite{Olivares03} 
and the non-locality due to the violation of Bell type inequality 
\cite{Nha04, Garcia-Patron04PRL, Garcia-Patron04, Olivares04}.

In this paper, 
we consider the photon subtraction scheme consisting of 
two on-off detectors. 
The scheme is applied to both single- and two-mode squeezed vacua 
including realistic parameters of possible imperfections. 
Special attention is paid to that the density operators conditioned 
by the on-off detector can be represented in terms of the sum of 
the three kinds of Gaussian states. 
Therefore photon subtracted non-Gaussian states 
may still include Gaussian nature more or less.

Based on this, we address the following two points. 
First, we discuss the validity of the second order variance as 
a measure of the performance of the photon subtracted state. 
Experimentally, the evaluation of the states by the second order variance of 
quantum fluctuations or correlations are easier and more accurate 
than the evaluation based on full reconstruction of the states such as 
quantum tomography. 
When two on-off detectors are applied to a single-mode squeezed vacuum, 
one can conditionally generate a plus-cat-like state which is often 
squeezed than the input. 
It is shown that in realistic squeezing regime, the squeezing 
of the photon subtracted state is higher than that of the input. 
We also show that photon subtractions from two-mode squeezed vacuum 
greatly enhance the second order quantum correlation of the output 
in practical parameter regime.

Second, we apply the photon subtracted entangled state to 
quantum dense coding \cite{Ban99,Braunstein00}. 
This is an entanglement-assisted coding to transmit 
classical information which can attain 
a larger capacity than that without entanglement \cite{Ralph02}. 
Therefore the increase of the capacity, more precisely, 
the mutual information for a specified measurement 
(e.g. the Bell measurement), 
can be an alternative operational measure 
for the photon subtracted entangled state. 
The mutual information is calculated by determining the channel 
matrix between the input signals and the measurement outcomes. 
This measure can be a stringent figure of merit for the system in 
the sense that the gain usually vanishes even with relatively small 
imperfections. 
This also specifies the asymptotic rate of transmission 
when the multiple use of the channels is considered. 
So it would be worth considering such kind of information theoretic 
measure to quantify the non-Gaussian state.

This paper is organized as follows: In Sec. \ref{single-NG-sec}, 
we discuss the measurement-induced non-Gaussian operation 
on single-mode squeezed vacuum state, and its equivalence 
to Schr\"{o}dinger cat-like state generation in Ref.~\cite{Dakna97}. In Sec. 
\ref{2mode-NG-sec}, we discuss the two-mode case. 
The generated non-Gaussian entangled state is applied to 
the dense coding scheme in Sec. \ref{coding}. 
In the Secs. \ref{POVM}, \ref{r-single-NG-sec}, and \ref{r-2mode-NG-sec}, 
we give analyses with consideration of practical parameters of 
imperfections. 
The last section 
\ref{discussion} is devoted to discussion and conclusion.

\section{Non-Gaussian operation on single-mode squeezed vacuum state} 
\label{single-NG-sec}

\begin{figure}
\centering 
\includegraphics[bb=55 55 420 445, width=.8\linewidth]{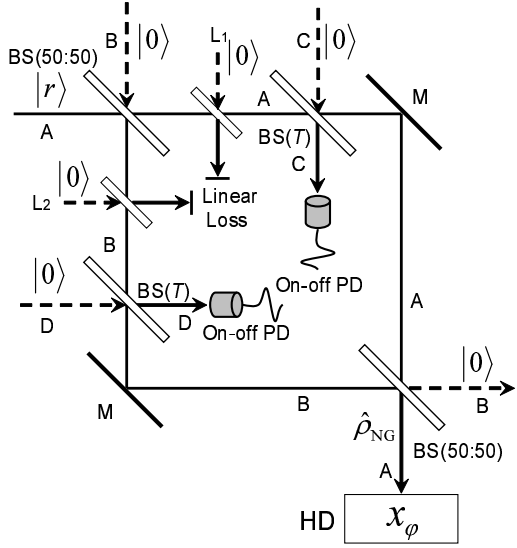}
\caption{\label{single-PS} Measurement-induced non-Gaussian operation 
on the single-mode squeezed vacuum state. 
BS, PD, M, and HD are beam splitter, photon detector, mirror, 
and homodyne detection, respectively. }
\end{figure}

We first consider the non-Gaussian operation on the single-mode 
squeezed vacuum state in the ideal situation. 
The schematic is shown in Fig. \ref{single-PS}. 
The target mode is denoted by path A. 
For the later extension to the two mode case, 
we consider another mode B whose initial state is the vacuum state. 
The initial state of mode A is the squeezed vacuum state 
\begin{equation}
|r\rangle _k =\hat{S} _k (r)|0\rangle , 
\end{equation}
where $\hat{S}_k (r)$ is the squeezing operator of mode $k$, 
\begin{equation}
\hat{S}_k (r)=\exp \left[ -\frac{r}{2} (\hat{a}_k ^{\dagger 2} -\hat{a}_k ^2 )
\right] , 
\end{equation}
and $r$ is the squeezing parameter, proportional 
to the second-order susceptibility and the thickness 
of the nonlinear optical crystal and the pump intensity. 
For this state, the uncertainty is reduced in terms of the $x-$ 
quadrature corresponding to $\varphi=0$ in the following 
quadrature operator 
\beq
\hat{x}_\varphi =\frac{1}{\sqrt{2}} \left( \hat{a}e^{-i\varphi } 
+\hat{a}^\dagger e^{i\varphi } \right).  
\label{QOP}
\eeq

The input squeezed state is divided into two modes A and B 
with a beam splitter of transmittance $\tau $.  
Mode C (D) is then tapped from mode A (B) 
with a beam splitter of high transmittance $T$. 
The resulting four-mode state is then 
\begin{equation}
|\psi _{\rm in} \rangle _{\rm ABCD} 
=\hat{V}_{\rm BD} (\theta )\hat{V}_{\rm AC} (\theta )
\hat{V}_{\rm AB} (\phi )|r\rangle _{\rm A} |0\rangle _{\rm BCD} , 
\end{equation}
where 
\begin{equation}
\hat{V}_{kl} (\theta )=\exp \left[ \theta (\hat{a}_k ^\dagger \hat{a}_l 
-\hat{a}_k \hat{a}_l ^\dagger )\right] 
\end{equation}
is the beam splitting operator, 
and the parameters $\phi $ and $\theta $ are 
related with the transmittances $\tau $ and $T$ as 
\begin{equation}
\tan \phi =\sqrt{\frac{1-\tau }{\tau } } , \ 
\tan \theta =\sqrt{\frac{1-T}{T} },   
\end{equation}
respectively. 
We consider a balanced interferometer $(\tau =0.5)$. 
Modes C and D are led to the on-off type photon detector. 
The probability operator valued measure (POVM) of on-off detector 
is described as 
\begin{equation}
\left\{ \hat{\Pi }^{(\textrm{on})} _k =\hat{1}_k -|0\rangle _k \langle 0|, 
\hat{\Pi }^{(\textrm{off})} _k =|0\rangle _k \langle 0| \right\}. 
\end{equation}
Simultaneous \textit{on} events in both modes C and D project the state 
over mode A and B into the state   
\begin{equation}
\hat{\rho } _{\rm out} =\frac{\textrm{Tr} _{\rm CD} 
\left[ |\psi _{\rm in} \rangle _{\rm (ABCD)} \langle \psi _{\rm in} |
\otimes \hat{\Pi }^{(\textrm{on})} _{\rm C} 
\otimes \hat{\Pi }^{(\textrm{on})} _{\rm D} \right] }
{P_{\rm det} } , \label{single-NG}
\end{equation}
where 
\begin{eqnarray}
P_{\rm det} &=&\textrm{Tr} _{\rm ABCD} 
\left[ |\psi _{\rm in} \rangle _{\rm (ABCD)} \langle \psi _{\rm in} |
\otimes \hat{\Pi }^{(\textrm{on})} _{\rm C} 
\otimes \hat{\Pi }^{(\textrm{on})} _{\rm D} \right] 
\nonumber \\
&=&1-2\sqrt{\frac{1-\lambda ^2 }{1-\lambda ^2 (\frac{1+T}{2})^2 } } 
+\sqrt{\frac{1-\lambda ^2 }{1-\lambda ^2 T^2 } } 
\end{eqnarray}
is the success probability of this \textit{on} event selection, 
and $\lambda \equiv \tanh r$. 
Finally, remaining modes A and B are recombined with another beam splitter 
of the transmittance $\tau (=0.5)$, being transformed as 
\begin{equation}
\hat{V}_{\rm AB} ^\dagger (\phi ) \hat{\rho } _{\rm out} 
\hat{V}_{\rm AB} (\phi ) =\hat{\rho }_{\rm NG} \otimes 
|0\rangle _{\rm B} \langle 0|. 
\end{equation}
The output state of mode B is the vacuum state due to the interference.

The success probability and the quality of the output non-Gaussian state 
can be controlled by the transmittance of the tapping beam splitter $T$.  
There is a trade-off between these two. 
The non-Gaussian property appears more strikingly for higher 
transmittance but with sacrifice of lower success probability. 
Hereafter, we set $T=0.9$.

The output non-Gaussian state in mode A is measured by the homodyne detection. 
The measurement basis is given by the the quadrature eigenstate 
\begin{eqnarray}
&&\hat{x}_\varphi |x_\varphi \rangle =x_\varphi |x_\varphi \rangle , \\ 
&&|x_\varphi \rangle =\frac{1}{\sqrt[4]{\pi } } 
\exp \left[ -\frac{1}{2}x_\varphi ^2 
+\sqrt{2}e^{i\varphi } x_\varphi \hat{a}^\dagger 
-\frac{1}{2} e^{2i\varphi } \hat{a} ^{\dagger 2} \right] |0\rangle . \nonumber \\
\label{QES} 
\end{eqnarray}

\begin{figure}
\centering 
\includegraphics[bb=60 50 560 750, angle=-90, width=.9\linewidth]{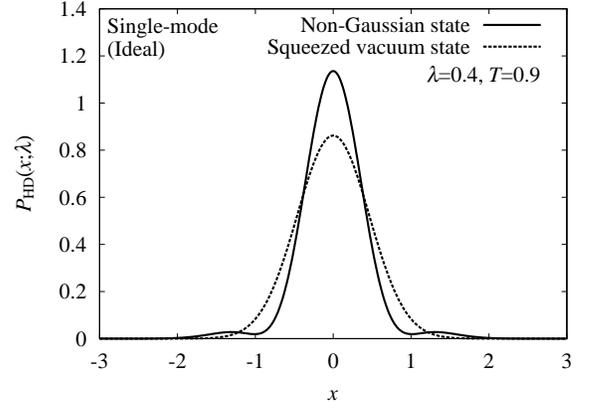}
\caption{\label{single-HD} Probability distribution of the single-mode 
non-Gaussian state (solid line, $\lambda =0.4$, $T=0.9$) 
and the squeezed vacuum state (dotted line, $\lambda =0.4$) 
for the phase parameter $\varphi =0$ with ideal setup. }
\end{figure}

The probability distribution is given by 
\begin{eqnarray}
P_{\rm HD} (x_\varphi ;\lambda )&=&\langle x_\varphi |\hat{\rho }_{\rm NG} 
|x_\varphi \rangle \nonumber \\
&=&P_{11} (x_\varphi ;\lambda )-P_{10} (x_\varphi ;\lambda ) \nonumber \\
&&\hspace{10mm} -P_{01} (x_\varphi ;\lambda )+P_{00} (x_\varphi ;\lambda ), 
\label{single-HD-eq}
\end{eqnarray}
where 
\begin{eqnarray}
\lefteqn{P_{ij} (x_\varphi ;\lambda )} \nonumber \\
&=&\frac{1}{\sqrt{\pi } P_{\rm det} } 
\sqrt{\frac{1-\lambda ^2 }
{(1-\lambda T)^2 -\lambda ^2 \gamma _{ij} ^2 
+4\lambda T\sin ^2 \varphi } } \nonumber \\
&&\times \exp \left[ -\frac{1-\lambda ^2 (T+\gamma _{ij} )^2 }
{(1-\lambda T)^2 -\lambda ^2 \gamma _{ij} ^2 +4\lambda T\sin ^2 \varphi } 
x_\varphi ^2 \right] , \nonumber \\
\end{eqnarray}
and $\gamma _{11}=R\equiv 1-T$, $\gamma _{10} =\gamma _{01} 
=R/2$, and $\gamma _{00}=0$. 
Thus the probability distribution consists of the four Gaussian terms. 
Fig. \ref{single-HD} shows 
the probability distribution of non-Gaussian state ($\lambda =0.4$) 
along $x_{\varphi =0} \equiv x$ axis (the solid line) 
and the one of the input squeezed vacuum state (the dotted line). 
The probability distribution of the output non-Gaussian state 
consists of the single main peak with two small side lobes. 
The probability distribution dominated by the single main peak may still 
be characterized by the variance 
\begin{eqnarray}
\lefteqn{V(\lambda )} \nonumber \\
&=&\int _{-\infty } ^\infty dx\ x^2 P_{\rm HD} (x;\lambda ) 
-\left\{ \int _{-\infty } ^\infty dx\ x P_{\rm HD} (x;\lambda ) \right\} ^2 
\nonumber \\
&=&V_{11} (\lambda )-V_{10} (\lambda )-V_{01} (\lambda )+V_{00} (\lambda ), 
\end{eqnarray}
where 
\begin{equation}
V_{ij} (\lambda )=\frac{\sqrt{1-\lambda ^2 } }{2P_{\rm det} } 
\frac{(1-\lambda T)^2 -\lambda ^2 \gamma _{ij} ^2 }
{\left[ 1-\lambda ^2 (T+\gamma _{ij} )^2 \right] ^{3/2} } 
\end{equation}
and $\gamma _{ij} $'s have been given above. 
\begin{figure}
\centering 
\includegraphics[bb=60 50 560 750, angle=-90, width=.9\linewidth]{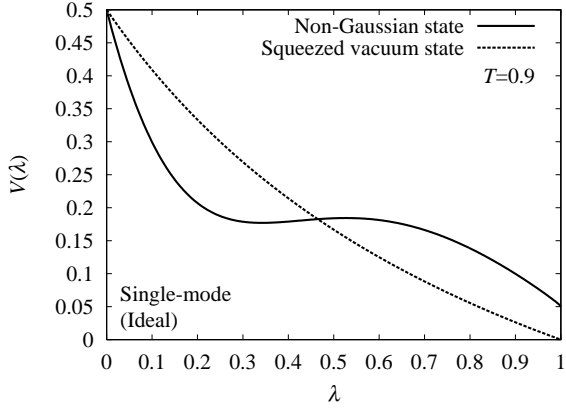}
\caption{\label{single-V} Variance of the single-mode non-Gaussian 
state (solid line, $T=0.9$) and the squeezed vacuum state 
(dotted line) with ideal setup. }
\end{figure}
In Fig. \ref{single-V}, 
the variances of the output non-Gaussian state and of the original input 
squeezed state are compared. 
As seen, in the range of $\lambda \lesssim 0.47$, 
the variance of the output non-Gaussian state is \textit{lower} 
than that of original squeezed state, 
which means that the squeezing degree is effectively enhanced.

\begin{figure}
\centering 
\includegraphics[bb=55 55 380 345, width=.8\linewidth]{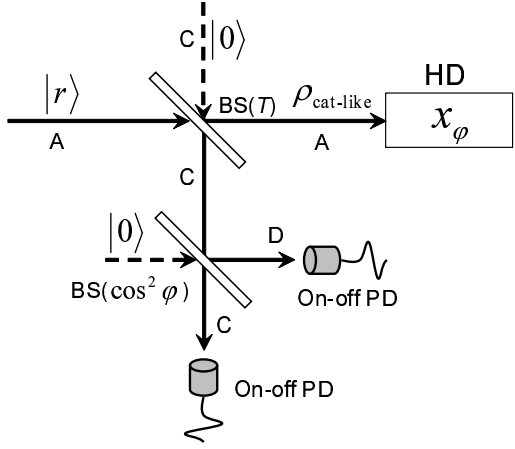}
\caption{\label{Dakna2} Schr\"{o}dinger cat-like state generation scheme 
in Ref. \cite{Dakna97}. BS, PD, and HD are beam splitter, photon detector, 
and homodyne detection, respectively. }
\end{figure}

Our interferometric non-Gaussian operation scheme is 
actually found to be equivalent to the Schr\"{o}dinger cat-like 
state generation proposed in Ref. \cite{Dakna97} 
(Fig. \ref{Dakna2}). 
This can directly be observed by 
the following equation 
\begin{eqnarray}
\lefteqn{\hat{V}_{\rm AB} ^\dagger (\phi ) 
|\psi _{\rm in} \rangle _{\rm ABCD} } 
\nonumber \\
&=&\hat{V}_{\rm CD} (\phi ) \hat{V}_{\rm AC} (\theta ) 
\hat{V}_{\rm BD} (\theta ) \hat{V}_{\rm CD} ^\dagger (\phi ) 
|r\rangle _{\rm A} |0\rangle _{\rm BCD} \nonumber \\
&=&\hat{V}_{\rm CD} (\phi ) \hat{V}_{\rm AC} (\theta ) 
|r\rangle _{\rm A} |0\rangle _{\rm BCD},  
\end{eqnarray}
by noting that input modes B, C, and D are all the vacuum state.

\begin{figure}
\centering 
\includegraphics[bb=100 80 480 680, angle=-90, width=.9\linewidth]{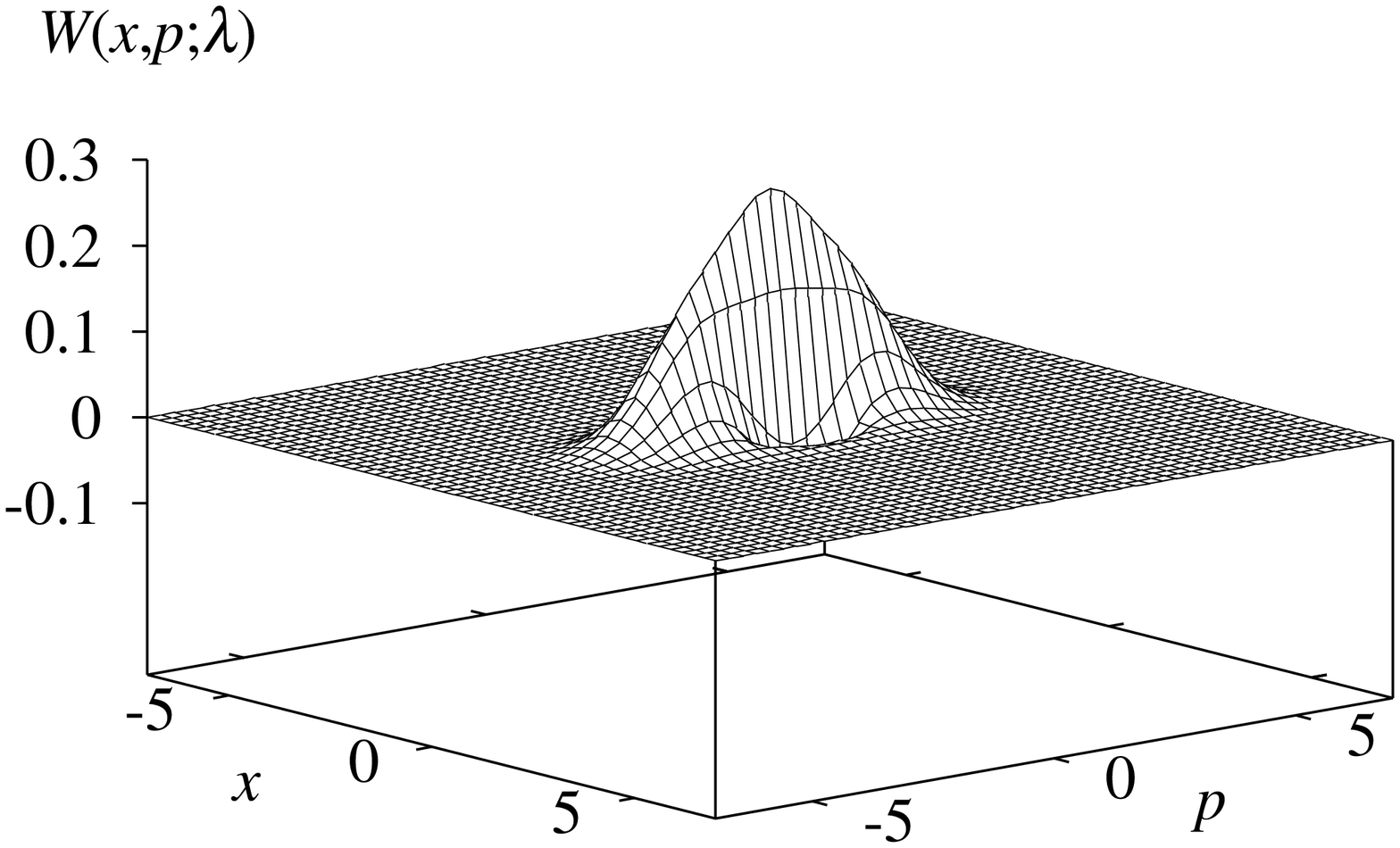}
\caption{\label{cat-like2-04} Wigner function of the single-mode non-Gaussian state 
($\lambda =0.4$, $T=0.9$) with ideal setup. }
\end{figure}

\begin{figure}
\centering 
\includegraphics[bb=100 80 480 680, angle=-90, width=.9\linewidth]{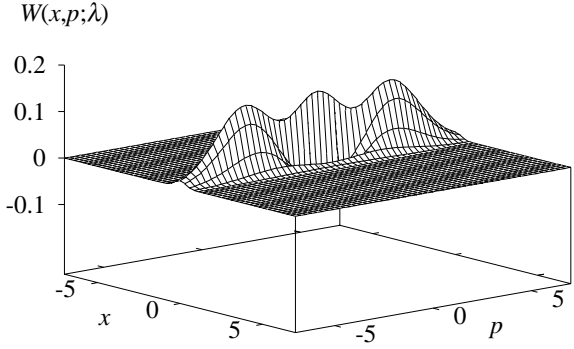}
\caption{\label{cat-like2-08} Wigner function of the single-mode non-Gaussian state 
($\lambda =0.8$, $T=0.9$) with ideal setup. }
\end{figure}

The Wigner function 
\begin{equation}
W(x,p)=\frac{1}{\pi } \int _{-\infty } ^\infty dy e^{-2ipy} \langle x-y|\rho |x+y\rangle 
\end{equation}
of the non-Gaussian state is obtained by 
\begin{eqnarray}
W_{\rm NG} (x,p;\lambda )&=&W_{11} (x,p;\lambda )-W_{10} (x,p;\lambda ) 
\nonumber \\
&&\hspace{2mm} -W_{01} (x,p;\lambda )+W_{00} (x,p;\lambda ), 
\end{eqnarray}
where 
\begin{eqnarray}
W_{ij} (x,p;\lambda )&=&\frac{1}{\pi P_{\rm det} } 
\sqrt{\frac{1-\lambda ^2 }{1-\lambda ^2 (T-\gamma _{ij} )^2 } } \nonumber \\
&&\hspace{3mm} \times \exp \left[ -\frac{1-\lambda ^2 (T+\gamma _{ij} )^2 }
{(1-\lambda T)^2 -\lambda ^2 \gamma _{ij} ^2 } x^2 \right] \nonumber \\
&&\hspace{6mm} \times \exp \left[ -\frac{(1-\lambda T)^2 -\lambda ^2 \gamma _{ij} ^2 }
{1-\lambda ^2 (T-\gamma _{ij} )^2 } p^2 \right] . \nonumber \\
\end{eqnarray}
This is illustrated for two $\lambda$'s  
in Figs. \ref{cat-like2-04} ($\lambda =0.4$) 
and \ref{cat-like2-08} ($\lambda =0.8$). 
For smaller $\lambda $, 
the Wigner function is closer to that of the squeezed vacuum state. 
For larger $\lambda $, on the other hand, 
the property of cat-like state becomes more remarkable. 
These figures will be compared later with the case including imperfections.

\begin{figure}
\centering 
\includegraphics[bb=60 59 555 740, angle=-90, width=.9\linewidth]{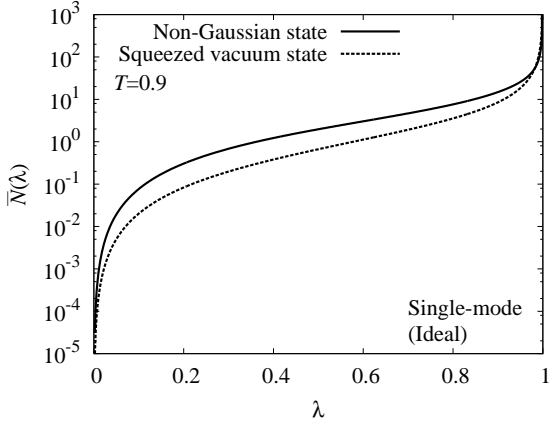}
\caption{\label{average_n-s} Average photon number of the single-mode 
non-Gaussian state (solid line, $T=0.9$) 
and the squeezed vacuum state (dotted line) with ideal setup. }
\end{figure}

Interestingly, the average photon number 
\begin{eqnarray}
\bar{N}(\lambda ) &=&\textrm{Tr} 
\left[ \hat{\rho } _{\rm NG} \otimes \hat{N} \right] \nonumber \\
&=&N_{11} (\lambda )-N_{10} (\lambda )-N_{01} (\lambda )+N_{00} (\lambda ) 
\end{eqnarray}
increases after \textit{conditional} non-Gaussian operation 
(Fig. \ref{average_n-s}), where $\hat{N} =\hat{a}^\dagger \hat{a} $ is 
the photon number operator, and 
\begin{equation}
N_{ij} (\lambda )=\frac{1}{P_{\rm det} } 
\sqrt{\frac{1-\lambda ^2 }{1-\lambda ^2 (T+\gamma _{ij} )^2 } } 
\frac{\lambda ^2 T(T+\gamma _{ij} )}
{1-\lambda ^2 (T+\gamma _{ij} )^2 } . 
\end{equation}

\section{Non-Gaussian operation on two-mode squeezed state} 
\label{2mode-NG-sec}

\begin{figure}
\centering 
\includegraphics[bb=55 55 500 445, width=.9\linewidth]{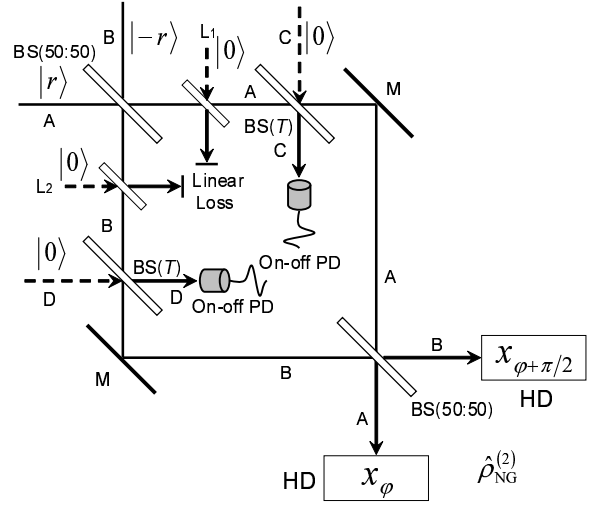}
\caption{\label{2mode-PS} Measurement-induced non-Gaussian operation 
on the two-mode squeezed vacuum state. 
BS, PD, M, and HD are beam splitter, photon detector, mirror, 
and homodyne detection, respectively. }
\end{figure}

Now we turn to the two-mode case (Fig. \ref{2mode-PS}) 
in the ideal situation, where the other squeezed vacuum state with 
the reduced uncertainty of the $p$ 
$(\equiv x_{\varphi +\frac{\pi }{2} } )$-quadrature 
is input into mode B 
instead of the vacuum state. 
The two-mode squeezed vacuum state 
\begin{eqnarray}
|r^{(2)} \rangle _{\rm AB} &=&\hat{V} _{\rm AB} \left(\frac{\pi }{4} \right) 
|r\rangle _{\rm A} |-r\rangle _{\rm B} \nonumber \\
&\equiv &\hat{S} ^{(2)} _{\rm AB} (-r)|0\rangle _{\rm AB} , 
\end{eqnarray}
is generated after the first 50:50 beam splitter, where 
\begin{equation}
\hat{S} ^{(2)} _{kl} (r)=\exp \left[ -r(\hat{a}_k ^\dagger \hat{a}_l ^\dagger 
-\hat{a}_k \hat{a}_l )\right] 
\end{equation}
is the two-mode squeezing operator. 
The operations after this is similar to the single-mode case. 
The state conditioned by the simultaneous $on$ results for modes C and D 
is 
\begin{equation}
\hat{\rho } _{\rm NG} ^{(2)} =\frac{\textrm{Tr} _{\rm CD} 
\left[ |\psi _{\rm in} ^{(2)} \rangle _{\rm (ABCD)} 
\langle \psi _{\rm in} ^{(2)} |
\otimes \hat{\Pi }^{(\textrm{on})} _{\rm C} \otimes \hat{\Pi }^{(\textrm{on})} _{\rm D} \right] }
{P_{\rm det} ^{(2)} } , \label{two-mode-NG}
\end{equation}
where 
\begin{equation}
|\psi _{\rm in} ^{(2)} \rangle _{\rm ABCD} 
=\hat{V}_{\rm AC} (\theta )\hat{V}_{\rm BD} (\theta ) |r^{(2)} \rangle _{\rm AB} 
|0\rangle _{\rm CD},  
\end{equation}
with the high transemittance $T=\cos ^2 \theta $, and 
\begin{eqnarray}
P_{\rm det} ^{(2)} &=&\textrm{Tr} _{\rm ABCD} 
\left[ |\psi _{\rm in} ^{(2)} \rangle _{\rm (ABCD)} 
\langle \psi _{\rm in} ^{(2)} |
\otimes \hat{\Pi }^{(\textrm{on})} _{\rm C} 
\otimes \hat{\Pi }^{(\textrm{on})} _{\rm D} \right] \nonumber \\
&=&\frac{\lambda ^2 (1-T)^2 (1+\lambda ^2 T)}{(1-\lambda ^2 T)(1-\lambda ^2 T^2 )}
\end{eqnarray}
is the success probability of the $on$ event selection.

\begin{figure}
\centering 
\includegraphics[bb=60 55 555 745, angle=-90, width=.9\linewidth]{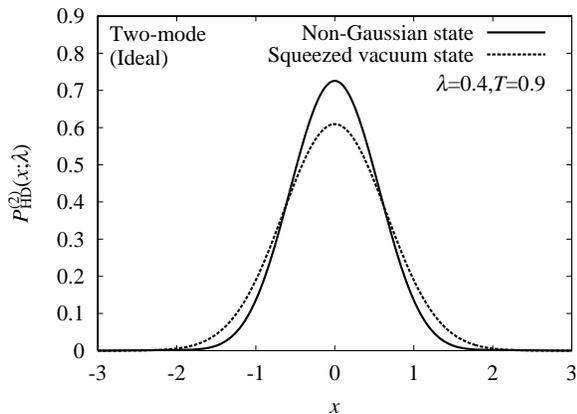}
\caption{\label{2mode-HD} Probability distribution of 
the two-mode non-Gaussian state (solid line, $\lambda =0.4$, $T=0.9$) 
and the squeezed vacuum state (dotted line, $\lambda =0.4$) 
for the phase parameter $\varphi =0$ on the mode A with ideal setup. 
Probability distribution for $\varphi =\pi /2 $ on the mode B gives the same result. }
\end{figure}

\begin{figure}
\centering 
\includegraphics[bb=60 55 555 745, angle=-90, width=.9\linewidth]{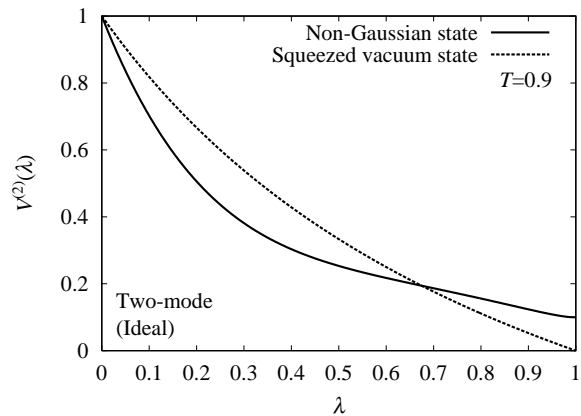}
\caption{\label{2mode-V} Variance of the two-mode non-Gaussian state 
(solid line, $T=0.9$) and the squeezed vacuum state (dotted line) 
with ideal setup. }
\end{figure}

The two modes of the conditional non-Gaussian entangled state are 
combined via the second 50:50 beam splitter, and are then 
measured by the two homodyne detectors, 
each of which measures the two orthogonal quadratures simultaneously. 
This chain of operations corresponds to the CV Bell measurement 
represented by 
\begin{equation}
|\Pi (x,p)\rangle _{kl} =\frac{1}{\sqrt{2\pi } } 
\int _{-\infty } ^\infty dy e^{ipy} |x+y\rangle _k |y\rangle _l . 
\label{Bell_basis}
\end{equation}
Therefore, the probability distribution of homodyne detection 
in phase space is 
\begin{eqnarray}
P_{\rm HD} ^{(2)} (x,p;\lambda )&=&
\langle \Pi (x,p)|\hat{\rho } _{\rm NG} ^{(2)} |\Pi (x,p)\rangle \nonumber \\
&=&P_{11} ^{(2)} (x,p;\lambda )-P_{10} ^{(2)} (x,p;\lambda ) \nonumber \\
&&\hspace{2mm} -P_{01} ^{(2)} (x,p;\lambda )+P_{00} ^{(2)} (x,p;\lambda ), 
\label{two-mode-HD}
\end{eqnarray}
where 
\begin{eqnarray}
\lefteqn{P_{ij} ^{(2)} (x,p;\lambda )} \nonumber \\
&=&\frac{1}{2\pi P_{\rm det} ^{(2)} } 
\frac{1-\lambda ^2 }{(1-\lambda T)^2 
-\lambda ^2 \gamma _i ^{(2)} \gamma _j ^{(2)} } \nonumber \\
&&\hspace{2mm} \times \exp \left[ -\frac{1
-\lambda ^2 (T+\gamma _1 ^{(2)} )(T+\gamma _2 ^{(2)} )}
{2\{ (1-\lambda T)^2 -\lambda ^2 \gamma _i ^{(2)} \gamma _j ^{(2)} \} } 
(x^2 +p^2 ) \right] , \nonumber \\
\end{eqnarray}
and $\gamma _1 ^{(2)} =R$, $\gamma _0 ^{(2)} =0$. We can see that 
Eq. (\ref{two-mode-HD}) is also the combination 
of four Gaussian terms, which is totally \textit{non-Gaussian} distribution. 
Fig. \ref{2mode-HD} shows the probability distribution 
$P_{\rm HD} ^{(2)} (x,p;\lambda )$ along $x$ axis for $\lambda =0.4$, 
after integrating out the variable $p$. 
In this two mode case, the two small side lobes cannot be seen, 
and the distribution is close to Gaussian. 
This distribution may be characterized by its variance, 
\begin{eqnarray}
\lefteqn{V^{(2)} (\lambda )} \nonumber \\
&=&\int _{-\infty } ^\infty dp \int _{-\infty } ^\infty 
dx\ x^2 P_{\rm HD} ^{(2)} (x,p;\lambda ) \nonumber \\
&&\hspace{20mm} 
-\left\{ \int _{-\infty } ^\infty dp \int _{-\infty } ^\infty 
dx\ x P_{\rm HD} ^{(2)} (x,p;\lambda ) \right\} ^2 
\nonumber \\
&=&V_{11} ^{(2)} (\lambda )-V_{10} ^{(2)} (\lambda )
-V_{01} ^{(2)} (\lambda )+V_{00} ^{(2)} (\lambda ), \label{2variance}
\end{eqnarray}
where 
\begin{equation}
V_{ij} ^{(2)} (\lambda )
=\frac{1}{P_{\rm det} ^{(2)} } 
\frac{(1-\lambda ^2 )\{ (1-\lambda T)^2 
-\lambda ^2 \gamma _i ^{(2)} \gamma _j ^{(2)} \} }
{[1-\lambda ^2 (T+\gamma _i ^{(2)} )(T+\gamma _j ^{(2)} )]^2 } . 
\end{equation}
This is shown in Fig. \ref{2mode-V} by the solid line. 
The dotted line corresponds to the case of the two-mode squeezed 
vacuum state. 
In $\lambda \lesssim 0.67$, the variance of two-mode non-Gaussian state 
is lower than that of two-mode squeezed vacuum state. 
Thus the squeezing degree 
is effectively enhanced in the two-mode case as well. 
The measurement on CV Bell basis (\ref{Bell_basis}) is one of 
the correlation between two modes, therefore we can consider 
the variance (\ref{2variance}) as a measure of quantum correlation 
of bipartite entangled pair.

\section{Application to dense coding scheme} \label{coding}

The entanglement measures for CV systems have been clarified for 
Gaussian states so far. 
But the measures for mixed non-Gaussian states are 
not clear yet. 
In this regards, it might be sensible to adopt some operational 
measures connected directly to CV protocols.  
In this section, we evaluate the property of the conditional 
non-Gaussian state in terms of the mutual information 
for the dense coding scheme proposed in \cite{Ban99,Braunstein00}. 
The result here is added to the ones based on the other kinds 
of operational measures, such as 
the improvement of the teleportation fidelity \cite{Olivares03} 
and the nonlocality due to the violation of Bell type inequality 
\cite{Nha04, Garcia-Patron04PRL, Garcia-Patron04, Olivares04}.

The variance reduction of the non-Gaussian states 
shown in the previous section implies the improvement of the 
signal-to-noise ratio, and hence the increase 
of the transmissible information. 
This can be measured by mutual information, which 
is calculated by evaluating the channel matrix between 
the input and output signals. 
This measure is known to be a stringent figure of merit 
for the system in the sense that the gain usually vanishes 
even with relatively small imperfections. 
Therefore it would be a nice criterion to see whether 
the information gain is observed or not 
by the non-Gaussian operation 
compared with the original two-mode squeezed vacuum state.

\begin{figure}
\centering 
\includegraphics[bb=55 55 570 405, width=.9\linewidth]{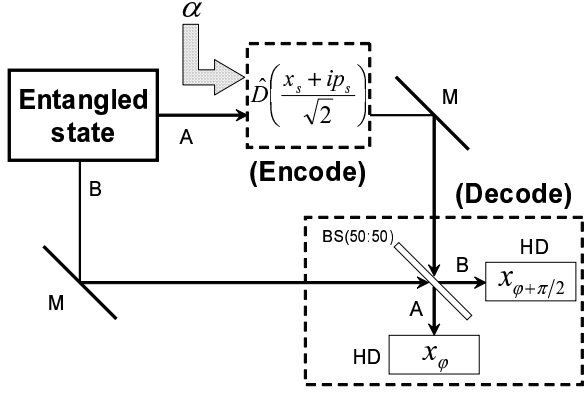}
\caption{\label{CV_coding} The setup of dense coding scheme with power of 
signal modulation $\alpha $. BS, M, and HD 
means beam splitter, mirror, and homodyne detection, respectively. }
\end{figure}

In Fig. \ref{CV_coding}, the setup of dense coding scheme is depicted. 
Alice encodes classical message $(x_s ,p_s)$ on one of the beam 
of the two-mode non-Gaussian entangled state 
by the displacement operation, 
\begin{eqnarray}
\hat{U} _{\rm A} (x_s ,p_s )&=&e^{-\frac{i}{2} x_s p_s }
\hat{D} \left( \frac{x_s +ip_s }{\sqrt{2} } \right) \nonumber \\
&=&e^{-ip_s \hat{x} _{\rm A} }e^{ix_s \hat{p}_{\rm A} } , 
\end{eqnarray}
where $\hat{D} (\alpha ) \equiv 
\exp (\alpha \hat{a}^\dagger -\alpha ^\ast \hat{a} )$. 
Now, we consider a simple model in which signals $x_s $ and 
$p_s $ are restricted to $\pm \sqrt{2} \alpha $ 
with equal prior probabilities,  
(quadrature phase shift keying) 
and thus a set of 2-bit information 
$\{ a_{00}=(x_s\equiv\sqrt{2}\alpha,p_s\equiv\sqrt{2}\alpha),\,
a_{01}=(\sqrt{2}\alpha,-\sqrt{2}\alpha),\,
a_{10}=(-\sqrt{2}\alpha,\sqrt{2}\alpha),\,
a_{11}=(-\sqrt{2}\alpha,-\sqrt{2}\alpha) \}$ 
is encoded. 
(The continuous encoding onto the non-Gaussian entangled state 
makes the calculation difficult, so we have here simplified to the 
4 valued discrete encoding. )

Bob attempts to decode 
classical message from Alice $(x_s ,p_s )$ 
by the CV Bell measurement described in Eq. (\ref{Bell_basis}). 
The distribution of the detection probability is given by 
\begin{eqnarray}
\lefteqn{\langle \Pi (x,p)|\hat{U} _{\rm A} (x_s ,p_s )
\hat{\rho } _{\rm NG} ^{(2)} \hat{U} _{\rm A} ^\dagger (x_s ,p_s ) 
|\Pi (x,p)\rangle } \nonumber \\
&&\hspace{30mm} =P_{\rm HD} ^{(2)} (x-x_s ,p-p_s ;\lambda ) \nonumber \\
&&\hspace{30mm} \equiv P(x,p|x_s ,p_s ), 
\end{eqnarray}
where we use the relation 
\begin{equation}
\hat{U}_{\rm A} ^\dagger (x_s ,p_s )|\Pi (x,p)\rangle 
=e^{-ip_s (x-x_s )} |\Pi (x-x_s ,p-p_s )\rangle . 
\end{equation}
Then Bob identifies the 2-bit classical messages according to 
the decision rule  
$\{ b_{00}=(x \ge 0,p \ge 0),\,
b_{01}=(x \ge 0,p < 0),\,
b_{10}=(x < 0,p \ge 0),\,
b_{11}=(x < 0,p < 0) \}$.

Now the 4-by-4 channel matrix is given by 
$\{ P(a_{mn}|b_{kl}) \}$ where the elements for ($m=k,n=l$) contribute 
to the success probability and otherwise, the error probability. 
For example, 
\begin{eqnarray}
\lefteqn{P(b_{00} |a_{00} )} \nonumber \\
&&=\int _0 ^\infty dx \int _0 ^\infty dp 
P(x,p|\sqrt{2}\alpha ,\sqrt{2}\alpha ) \nonumber \\
&&=\sum _{i,j=0} ^1 \frac{1}{4P_{\rm det} ^{(2)} } 
\frac{1-\lambda ^2 }{1-\lambda ^2 (T+\gamma ^{(2)} _i )(T+\gamma ^{(2)} _j )}
\nonumber \\
&&\hspace{30mm} 
\times \left( 1+\textrm{erf}\left[ \Omega _{ij} (\lambda ) \alpha \right] 
\right) ^2 \label{ch-mtrx-component}
\end{eqnarray}
where 
\begin{equation}
\Omega _{ij} (\lambda )
=\sqrt{
\frac{1-\lambda ^2 (T+\gamma ^{(2)} _i )(T+\gamma ^{(2)} _j )}
{(1-\lambda T)^2 -\lambda ^2 \gamma ^{(2)} _i \gamma ^{(2)} _j } } 
\end{equation}
and 
\begin{equation}
\textrm{erf}\ \!(x)=\frac{2}{\sqrt{\pi } } \int _0 ^x dte^{-t^2 } 
\end{equation}
is the error function. 
Other components can be obtained similarly, 
\begin{eqnarray}
\lefteqn{P(b_{mn} |a_{kl} )} \nonumber \\
&&=\sum _{i,j=0} ^1 \frac{1}{4P_{\rm det} ^{(2)} } 
\frac{1-\lambda ^2 }{1-\lambda ^2 (T+\gamma ^{(2)} _i )(T+\gamma ^{(2)} _j )}
\nonumber \\
&&\hspace{10mm} 
\times \left( 1+(-1)^{m-k} \textrm{erf}\left[ \Omega _{ij} (\lambda ) \alpha \right] 
\right) \nonumber \\
&&\hspace{20mm} 
\times \left( 1+(-1)^{n-l} \textrm{erf}\left[ \Omega _{ij} (\lambda ) \alpha \right] 
\right) . 
\end{eqnarray}
The mutual information is calculated by the above channel matrix as 
\begin{eqnarray}
\lefteqn{I(\textrm{A};\textrm{B})} \nonumber \\
&&=\sum _{k,l,m,n} P(a_{kl} )P(b_{mn} |a_{kl} ) \nonumber \\
&&\hspace{10mm} \times \log _2 \left[ \frac{P(b_{mn} |a_{kl} )}
{\displaystyle \sum _{k,l} P(a_{kl} )P(b_{mn} |a_{kl} )} \right] \nonumber \\
&&\rightarrow 2\ [\textrm{bit}]\quad \textrm{as}\ \lambda \rightarrow 1 , 
\end{eqnarray}
where $P(a_{kl} )=1/4$ is the prior probability.

From the viewpoint of communication efficiency, 
we should optimize the mutual information under the power constraint 
condition that the total energy of initial squeezing power 
and encoding displacement power is constant. We are, however, 
interested in how to quantify the improvement brought by the entanglement 
enhancement due to the non-Gaussian operation for a given input. 
What should be fixed is then the degree of entanglement 
of the input state, equivalently the squeezing degree $\lambda $ 
for the input. We intend to compare the mutual informations 
between the cases with and without the non-Gaussian operation.

\begin{figure}
\centering 
\includegraphics[bb=60 55 555 745, angle=-90, width=.9\linewidth]{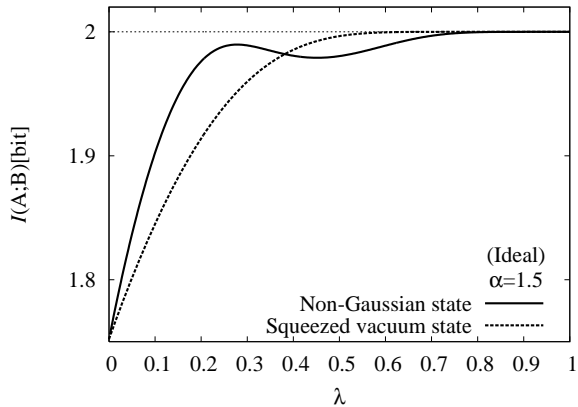}
\caption{\label{mutual_info_15} Mutual information via the dense coding channel 
($\alpha =1.5$) of the non-Gaussian state (solid line, $T=0.9$) 
and the squeezed vacuum state (dotted line) with ideal setup. }
\end{figure}

\begin{figure}
\centering 
\includegraphics[bb=60 55 555 745, angle=-90, width=.9\linewidth]{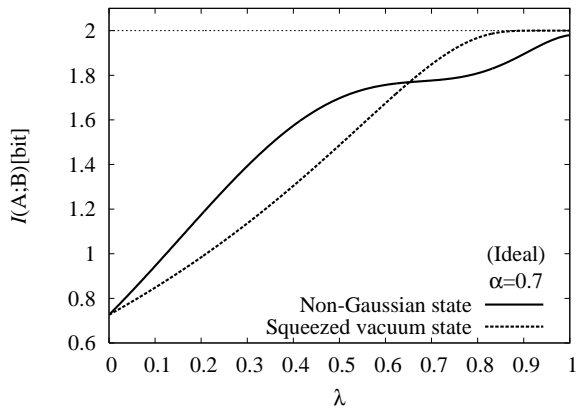}
\caption{\label{mutual_info_07} Mutual information via the dense coding channel 
($\alpha =0.7$) of the non-Gaussian state (solid line, $T=0.9$) 
and the squeezed vacuum state (dotted line) with ideal setup.  }
\end{figure}

In Figs. \ref{mutual_info_15} and \ref{mutual_info_07}, 
the mutual information of $\alpha =1.5$ and $\alpha =0.7$ 
are indicated respectively, 
where the original squeezed vacuum state cases 
are also indicated by the dotted line. 
We can see that, in the range of $\lambda \lesssim 0.38$ 
and $\lambda \lesssim 0.65$, in Fig. \ref{mutual_info_15} 
and \ref{mutual_info_07}, respectively, 
the mutual information of two-mode non-Gaussian state is larger 
than the original two-mode squeezed vacuum state case. 
This can naturally be regarded as the gain assisted by the 
enhanced entanglement of the non-Gaussian entangled state. 
Via the probabilistic photon subtraction operation, 
similar to the single-mode case, the average photon number 
of output two-mode non-Gaussian state increases compared with 
input squeezed vacuum state, which brings effective enhancement 
of entanglement.

\section{POVM of practical photon detector model} \label{POVM}

From this section, we consider possible imperfections, 
including the linear loss in optical paths, finite quantum efficiency 
and nonzero dark counts of the on-off detector. 
The imperfect photon detector may be modeled by the following 
POVM element \cite{Barnett98},  
\begin{equation}
\hat{\Pi } (m;\eta )=\sum _{n=m} ^\infty \binom{n}{m} 
\eta ^m (1-\eta )^{n-m} |n\rangle \langle n|, 
\end{equation}
where $n$ photons are converted to $m\leq n$ with the imperfect 
quantum efficiency $\eta $, 
and 
\begin{equation}
\binom{n}{m} =\frac{n!}{(n-m)!m!} 
\end{equation}
is the binomial coefficient. 
This element is considered as the sequence of two devices;
beam splitter of transmittance $\eta $ (linear loss) 
and perfect $n$ photon counting detector, where
$n-m$ photons are removed as the linear loss. 
As a whole, $m$ photons are detected. 

Further, assume that the net count 
of photon detector 
is $N(\geq m)$, where $N-m$ photons are due to dark count, 
following Poisson distribution. 
With average dark count $\nu $, 
the POVM of detector counting $N$ photons is 
\begin{eqnarray}
\lefteqn{ \hat{\Pi } (N,\eta ,\nu )} \nonumber \\
&&=\sum _{m=0} ^N e^{-\nu } 
\frac{\nu ^{N-m} }{(N-m)!} \sum _{n=m} ^\infty \binom{n}{m} 
\eta ^m (1-\eta )^{n-m} |n\rangle \langle n|, \nonumber \\
&&\label{POVM-N}
\end{eqnarray}
which satisfies completeness relation, 
\begin{equation}
\sum _{N=0} ^\infty \hat{\Pi } (N,\eta ,\nu ) =\hat{1}. \label{completeness}
\end{equation} 
With Eqs. (\ref{POVM-N}) and (\ref{completeness}), we obtain a set of POVM's 
of practical on-off type photon detector, 
\begin{equation}
\left\{ \hat{\Pi } ^{(\textrm{off})} (\eta ,\nu ),
\hat{\Pi } ^{(\textrm{on})} (\eta ,\nu ) \right\}, 
\end{equation}
where 
\begin{eqnarray}
\hat{\Pi } ^{(\textrm{off})} (\eta ,\nu ) &=&\hat{\Pi } (0,\eta ,\nu ) \nonumber \\
&=&e^{-\nu } \sum _{n=0} ^\infty 
(1-\eta )^n |n\rangle \langle n| , \\
\hat{\Pi } ^{(\textrm{on})} (\eta ,\nu ) &=&\sum _{N=1} ^\infty \hat{\Pi } (N,\eta ,\nu ) 
\nonumber \\
&=&\hat{1}-\hat{\Pi } ^{(\textrm{off})} (\eta ,\nu ) . 
\end{eqnarray}

\section{Analysis with practical parameters in single-mode scheme} 
\label{r-single-NG-sec}

Let us first go with the single-mode non-Gaussian operation 
(Fig. \ref{single-PS}). 
The quantum efficiency of the homodyne detector may be considered 
as approximately the unity, which is the case of the time limited 
signals prepared by chopping the continuous wave field generated 
from the cavity enhanced optical parametric process. 
We assume that the linear loss occurs between the first and 
second beam splitters, 
which may be attributed to the loss on beam splitters, 
mirrors, nonlinear crystals. 
Such a decohered squeezed vacuum state is transformed 
into the non-Gaussian state 
by the imperfect on-off type photon detectors modeled above.

The linear loss in optical paths can be described with a beam splitter 
of transmittance $T_L $, 
\begin{eqnarray}
\lefteqn{|\psi _{\rm in} ' \rangle _{\rm ABCDL_1 L_2} } \nonumber \\
&&=\hat{V}_{\rm BD} (\theta ) \hat{V} _{\rm AC} (\theta )
\hat{V} _{\rm BL_2 }(\xi ) \hat{V} _{\rm AL_1 } (\xi ) \nonumber \\
&&\hspace{20mm} \times \hat{V} _{\rm AB} \left( \frac{\pi }{4} \right) |r\rangle _{\rm A} 
|0\rangle _{\rm BCDL_1 L_2} 
\end{eqnarray}
where $L_1 $ and $L_2 $ are assigned for loss channels and 
\begin{equation}
\tan \xi =\sqrt{\frac{1-T_L }{T_L } } .
\end{equation}
Modes of $L_1 $ and $L_2 $ are traced out,  
and mode C (D) is tapped from mode A (B) by a beam splitter 
of high transmittance $T$ respectively. Therefore input state is 
\begin{equation}
\hat{\varrho } _{\rm in} =\textrm{Tr} _{\rm L_1 L_2 } 
\left[ |\psi ' _{\rm in} \rangle _{\rm (ABCDL_1 L_2 )} 
\langle \psi ' _{\rm in}| \right] . 
\end{equation}
Then, simultaneous \textit{on} events on modes C and D 
are selected conditionally, 
\begin{equation}
\hat{\varrho }_{\rm out} =\frac{\textrm{Tr} _{\rm CD} \left[ 
\hat{\varrho } _{\rm in} 
\otimes \hat{\Pi }^{(\textrm{on})} _{\rm C} (\eta ,\nu ) 
\otimes \hat{\Pi }^{(\textrm{on})} _{\rm D} (\eta ,\nu ) 
\right] }{{\cal P} _{\rm det} } , 
\end{equation}
where 
\begin{eqnarray}
{\cal P}_{\rm det} &=&\textrm{Tr} _{\rm ABCD} \left[ 
\hat{\varrho } _{\rm in} 
\otimes \hat{\Pi }^{(\textrm{on})} _{\rm C} (\eta ,\nu ) 
\otimes \hat{\Pi }^{(\textrm{on})} _{\rm D} (\eta ,\nu ) 
\right] \nonumber \\
&=&1-2e^{-\nu } \sqrt{\frac{1-\lambda ^2 }
{1-\lambda ^2 (T_L T+R_L +\frac{2-\eta }{2}T_L R)^2 }} 
\nonumber \\
& &+e^{-2\nu } \sqrt{\frac{1-\lambda ^2 }
{1-\lambda ^2 \{ T_L T+R_L +(1-\eta )T_L R\} ^2 }} \nonumber \\
\end{eqnarray}
is the success probability of event selection ($R_L \equiv 1-T_L $). 
Then we recombine modes A and B with another balanced beam splitter 
and mode A is measured with homodyne detection (mode B is vacuum state), 
\begin{eqnarray}
{\cal P}_{\rm HD} (x_\varphi ;\lambda )&=&
\langle x_\varphi |\hat{\varrho }_{\rm NG} |x_\varphi \rangle \nonumber \\
&=&{\cal P}_{11} (x_\varphi ;\lambda )
-e^{-\nu } {\cal P}_{10} (x_\varphi ;\lambda ) \nonumber \\
&&\hspace{2mm} -e^{-\nu } {\cal P}_{01} (x_\varphi ;\lambda )
+e^{-2\nu } {\cal P}_{00} (x_\varphi ;\lambda ), \nonumber \\
\end{eqnarray}
where 
\begin{widetext}
\begin{eqnarray}
\hat{\varrho }_{\rm NG} &=&
\hat{V}_{\rm AB} ^\dagger \left( \frac{\pi }{4} \right) 
\hat{\varrho }_{\rm out} \hat{V}_{\rm AB} \left( \frac{\pi }{4} \right) , \\
{\cal P}_{ij} (x_\varphi ;\lambda )&=&\frac{1}{\sqrt{\pi } {\cal P}_{\rm det} } 
\sqrt{\frac{1-\lambda ^2 }{(1-\lambda T_L T)^2 -\lambda ^2 (R_L +\gamma ' _{ij} )^2 
+4\lambda T_L T\sin ^2 \varphi }} \nonumber \\
&&\hspace{30mm} \times 
\exp \left[ -\frac{1-\lambda ^2 (T_L T+R_L +\gamma ' _{ij} )^2 }
{(1-\lambda T_L T)^2 -\lambda ^2 (R_L +\gamma ' _{ij} )^2 
+4\lambda T_L T\sin ^2 \varphi  } x_\varphi ^2 \right] , 
\end{eqnarray}
\end{widetext}
and $\gamma ' _{11} =T_L R$, $\gamma ' _{10} =\gamma ' _{01} =(2-\eta )T_L R/2$, 
and $\gamma ' _{00} =(1-\eta )T_L R$. With above results, we can obtain 
the variance of non-Gaussian state along $x$ axis ($\varphi =0$), 
\begin{eqnarray}
\lefteqn{{\cal V} (\lambda )} \nonumber \\
&=&\int _{-\infty } ^\infty dx\ 
x^2 {\cal P}_{\rm HD} (x;\lambda ) 
-\left\{ \int _{-\infty } ^\infty dx\ x{\cal P} _{\rm HD} (x;\lambda ) \right\} ^2 
\nonumber \\
&=&{\cal V}_{11} (\lambda ) -e^{-\nu } {\cal V}_{10} (\lambda ) 
-e^{-\nu } {\cal V}_{01} (\lambda ) +e^{-2\nu } {\cal V}_{00} (\lambda ), 
\nonumber \\
\end{eqnarray}
where 
\begin{eqnarray}
{\cal V} _{ij} (\lambda )&=&\frac{1}{2{\cal P}_{\rm det} } 
\sqrt{\frac{1-\lambda ^2 }{1-\lambda ^2 (T_L T+R_L +\gamma ' _{ij} )^2 }} 
\nonumber \\
&&\hspace{5mm} \times 
\frac{(1-\lambda T_L T)^2 -\lambda ^2 (R_L +\gamma ' _{ij} )^2 }
{1-\lambda ^2 (T_L T+R_L +\gamma ' _{ij} )^2 } . 
\end{eqnarray}

We assume the total linear loss 25\% ($T_L =0.75$), 
and the on-off type detector of quantum efficiency $\eta =0.6$ 
and the dark count rate 10000 [counts/sec]. At present, lower dark count rate 
$\sim 100$ [counts/sec] is achievable in laboratory, 
however, we consider the rather 
large value to see the effect more clearly. 
When the gating time of photon detector 
is $10^{-7} $ [sec], the net dark count in a single event is $\nu =10^{-3}$, 
which is large enough to affect the output state.

\begin{figure}
\centering 
\includegraphics[bb=60 55 555 745, angle=-90, width=.9\linewidth]{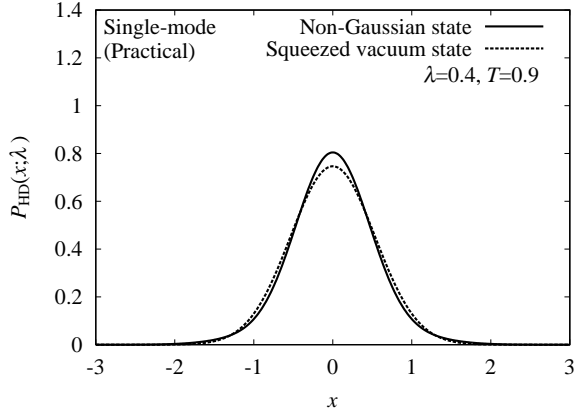}
\caption{\label{r-single-HD} Probability distribution 
of the single-mode non-Gaussian state (solid line, $\lambda =0.4$, $T=0.9$) 
and the squeezed vacuum state (dotted line, $\lambda =0.4$) 
for the phase parameter $\varphi =0$ with practical setup ($T_L =0.75$, 
$\eta =0.6$, $\nu =10^{-3} $). }
\end{figure}

\begin{figure}
\centering 
\includegraphics[bb=60 55 555 745, angle=-90, width=.9\linewidth]{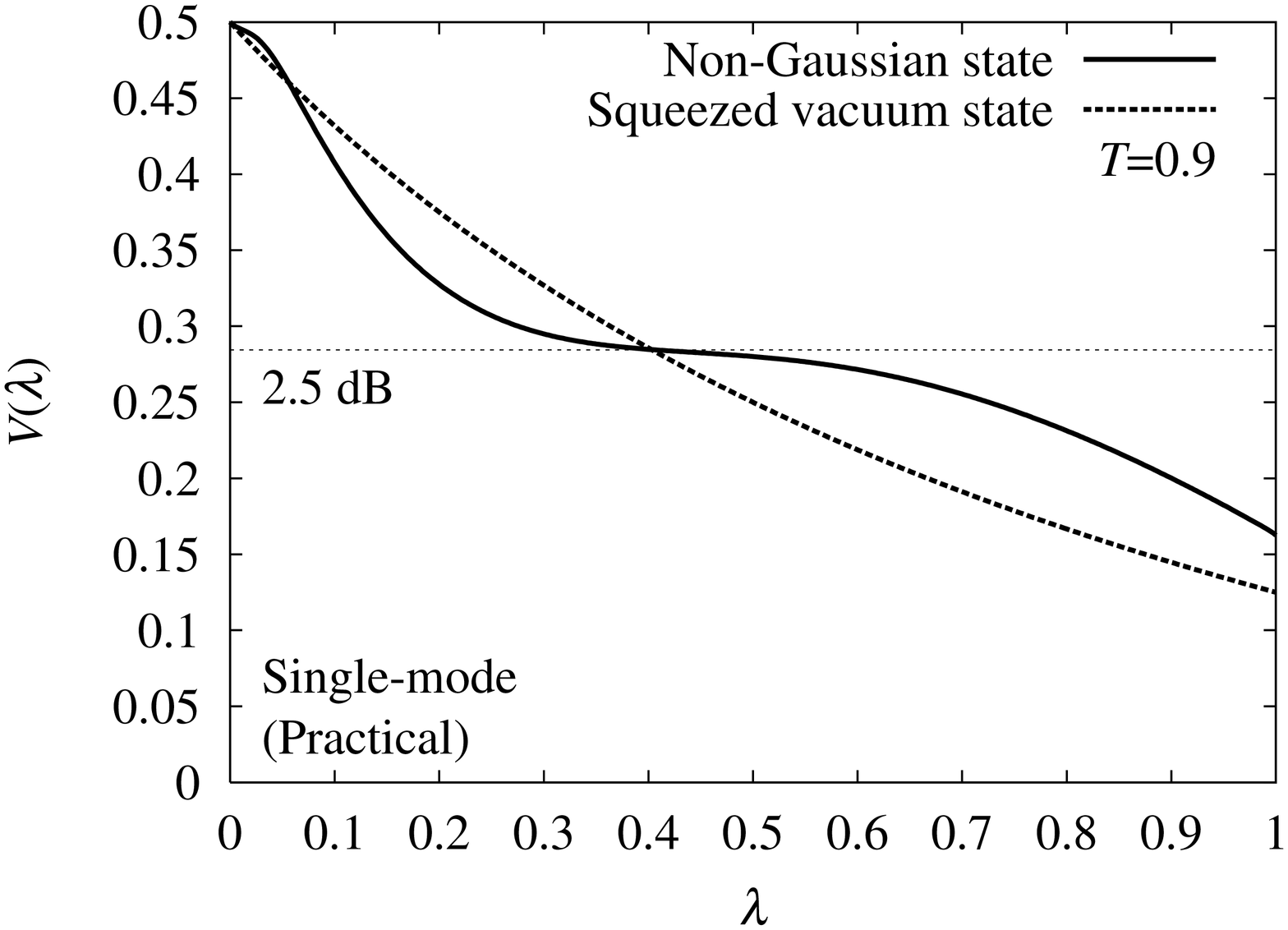}
\caption{\label{r-single-V} Variance 
of the single-mode non-Gaussian state (solid line, $T=0.9$) 
and the squeezed vacuum state (dotted line) with practical setup 
($T_L =0.75$, $\eta =0.6$, $\nu =10^{-3} $). }
\end{figure}

Figure \ref{r-single-HD} shows the probability distribution 
of non-Gaussian state ($\lambda =0.4$) along $x$ axis (solid line) 
and the one of the input squeezed vacuum state (dotted line), 
where the two side lobes seen in Fig. \ref{single-HD} 
disappear due to imperfections. 
In Fig. \ref{r-single-V}, the variances of the output non-Gaussian state 
and of the original input squeezed state are compared. 
The reduction of the variance due to the non-Gaussian operation 
is seen for $\lambda \lesssim 0.4$. At this cross point, 
the variance shows 2.5 dB (in terms of 
$-10\log _{10} \left[ {\cal V}(\lambda )/{\cal V}(0) \right] 
[\textrm{dB}]$) below the shot noise level. 
The range of the variance suppression becomes narrower 
than the ideal case where $\lambda \lesssim 0.47$ (Fig. \ref{single-V}). 
This is mainly due to the dark counts. 
The linear loss spoils the degree of squeezing, 
as the increase of the overall variance level. 
The imperfect quantum efficiency 
reduces success probability ${\cal P}_{\rm det} $. 

\begin{figure}
\centering 
\includegraphics[bb=100 80 470 680, angle=-90, width=.9\linewidth]{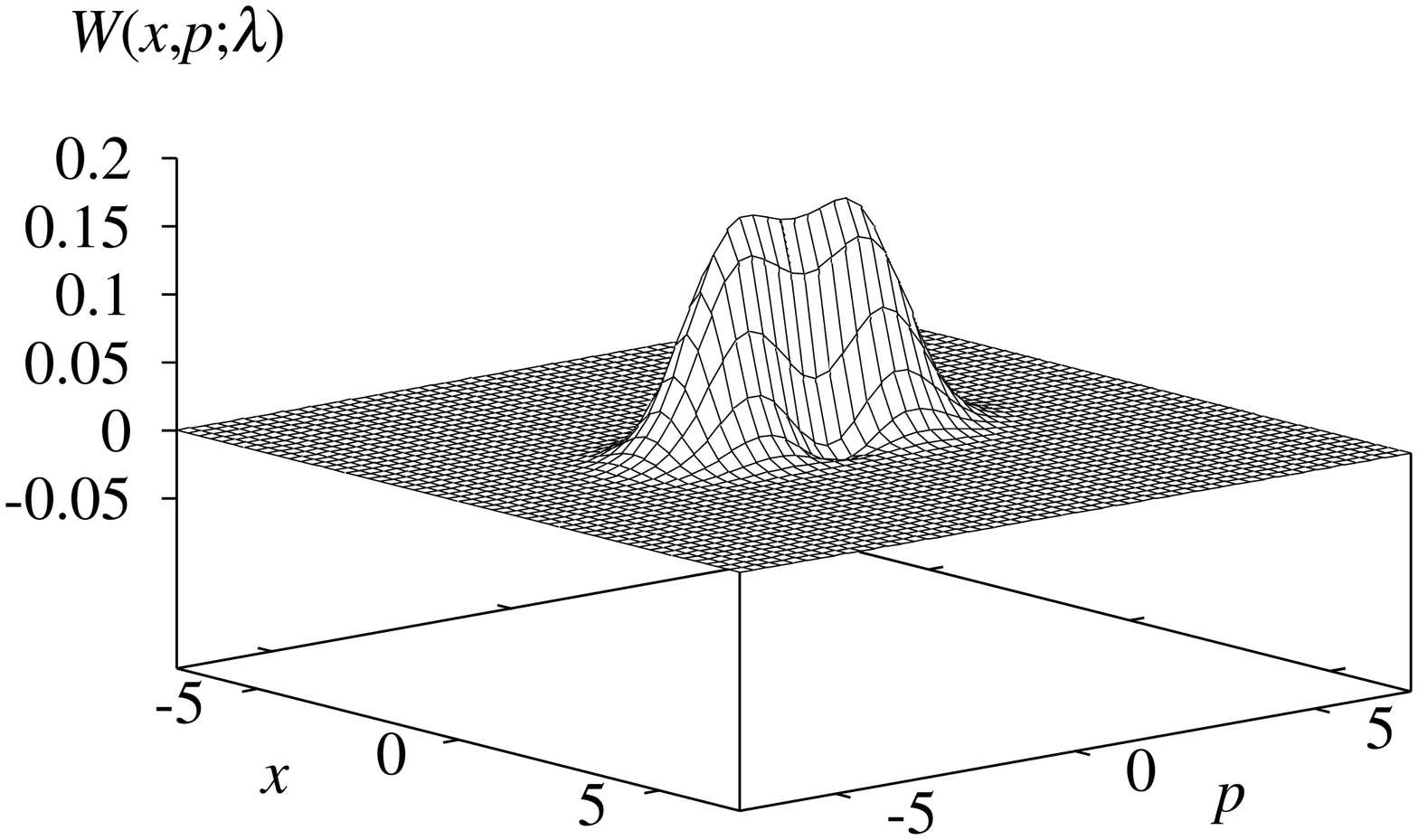}
\caption{\label{r-cat-like2-04} Wigner function of the single-mode non-Gaussian state 
($\lambda =0.4$, $T=0.9$) with practical setup ($T_L =0.75$, $\eta =0.6$, $\nu =10^{-3} $). }
\end{figure}

\begin{figure}
\centering 
\includegraphics[bb=100 80 470 680, angle=-90, width=.9\linewidth]{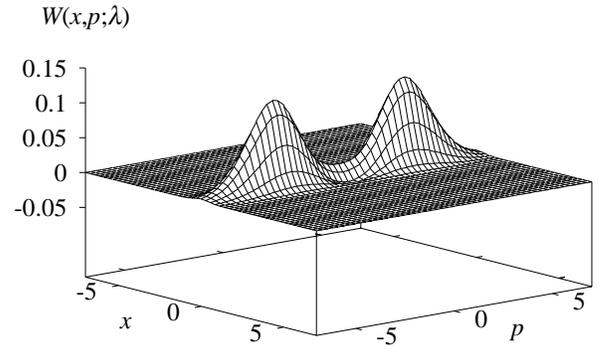}
\caption{\label{r-cat-like2-08} Wigner function of the single-mode non-Gaussian state 
($\lambda =0.8$, $T=0.9$) with practical setup ($T_L =0.75$, $\eta =0.6$, $\nu =10^{-3} $). }
\end{figure}

The Wigner function of the non-Gaussian state in the imperfect setup 
is calculated as 
\begin{eqnarray}
\lefteqn{{\cal W}_{\rm NG} (x,p;\lambda )} \nonumber \\
&=&{\cal W}_{11} (x,p;\lambda )
-e^{-\nu } {\cal W}_{10} (x,p;\lambda ) \nonumber \\
&&\hspace{10mm} 
-e^{-\nu } {\cal W}_{01} (x,p;\lambda )+e^{-2\nu } {\cal W}_{00} (x,p;\lambda ), 
\end{eqnarray}
where 
\begin{eqnarray}
\lefteqn{{\cal W}_{ij} (x,p;\lambda )} \nonumber \\
&=&\frac{1}{\pi {\cal P}_{\rm det} } 
\sqrt{\frac{1-\lambda ^2 }{1-\lambda ^2 (T_L T-R_L -\gamma ' _{ij} )^2 } } 
\nonumber \\
&&\hspace{1mm} \times \exp \left[ -\frac{1
-\lambda ^2 (T_L T+R_L +\gamma ' _{ij} )^2 }
{(1-\lambda T_L T)^2 -\lambda ^2 (R_L +\gamma ' _{ij} )^2 } 
x^2 \right] \nonumber \\
&&\hspace{3mm} \times \exp \left[ -\frac{(1-\lambda T_L T)^2 
-\lambda ^2 (R_L +\gamma ' _{ij} )^2 }
{1-\lambda ^2 (T_L T-R_L -\gamma ' _{ij} )^2 } p^2 \right] . 
\end{eqnarray}
The Wigner functions for $\lambda =0.4$ and $\lambda =0.8$ are illustrated in 
Figs. \ref{r-cat-like2-04} and \ref{r-cat-like2-08}, respectively. 
The former is still similar to the squeezed state, 
while the latter is close to decohered plus-cat state, compared with 
Figs. \ref{cat-like2-04} and \ref{cat-like2-08}. 

\section{Analysis with practical parameters in two-mode scheme}
\label{r-2mode-NG-sec}

\begin{figure}
\centering 
\includegraphics[bb=60 55 555 745, angle=-90, width=.9\linewidth]{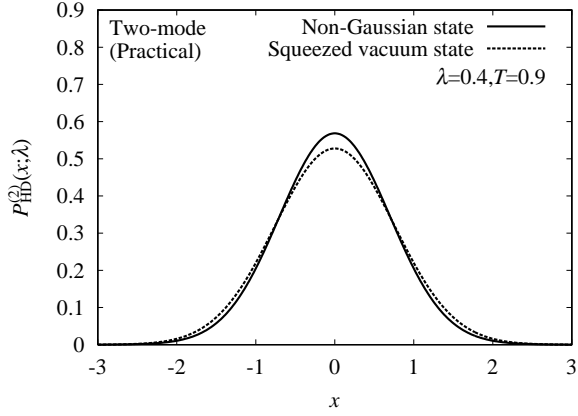}
\caption{\label{r-2mode-HD} Probability distribution 
of the two-mode non-Gaussian state (solid line, $\lambda =0.4$, $T=0.9$) 
and the squeezed vacuum state (dotted line, 
$\lambda =0.4$) for the phase parameter $\varphi =0$ 
on the mode A with practical setup ($T_L =0.75$, $\eta =0.6$, $\nu =10^{-3} $). 
Probability distribution for $\varphi =\pi /2 $ on the mode B gives the same result. }
\end{figure}

\begin{figure}
\centering 
\includegraphics[bb=60 55 555 745, angle=-90, width=.9\linewidth]{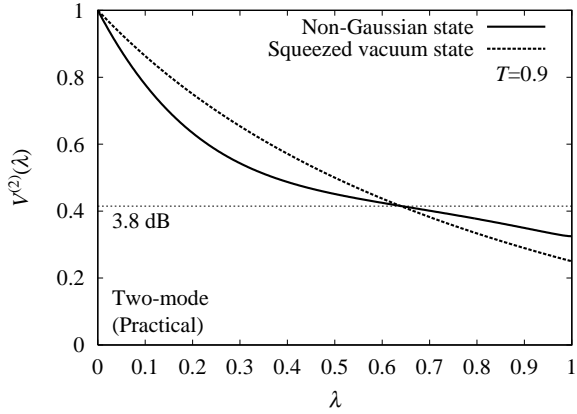}
\caption{\label{r-2mode-V} Variance of the two-mode non-Gaussian state 
(solid line, $T=0.9$) and the squeezed vacuum state (dotted line) 
with practical setup ($T_L =0.75$, $\eta =0.6$, $\nu =10^{-3} $). }
\end{figure}

We turn to the analysis of two-mode scheme with practical parameters 
(Fig. \ref{2mode-PS}). 
Similar to the single-mode case, we assume that the linear loss occurs 
between the first and second beam splitters, and model the effect 
by inserting a beam splitter of transmittance $T_L$ into each arm. 
The resulting state is then 
\begin{equation}
\hat{\varrho }_{\rm in} ^{(2)} 
=\textrm{Tr} _{\rm L_1 L_2 } \left[ 
|\psi ^{\prime (2)} _{\rm in} \rangle _{\rm (ABCDL_1 L_2 )} 
\langle \psi ^{\prime (2)} _{\rm in} |\right] , \label{r-2mode-NG} 
\end{equation}
where 
\begin{eqnarray}
\lefteqn{|\psi ^{\prime (2)} _{\rm in} \rangle _{\rm ABCDL_1 L_2 } } 
\nonumber \\
&=&\hat{V}_{\rm BD} (\theta ) \hat{V} _{\rm AC} (\theta ) 
\hat{V}_{\rm BL_2 } (\xi )\hat{V}_{\rm AL_1 } (\xi )
\hat{V} _{\rm AB} \left( \frac{\pi }{4} \right) \nonumber \\
&&\hspace{25mm} \times |r\rangle _{\rm A} |-r\rangle _{\rm B} |0\rangle _{\rm CDL_1 L_2 } . 
\end{eqnarray}
The conditional state selected by simultaneous \textit{on} events 
on modes C and D is then, 
\begin{equation}
\hat{\varrho }_{\rm NG} ^{(2)} 
=\frac{\textrm{Tr} _{\rm CD} \left[ \hat{\varrho }_{\rm in} ^{(2)} 
\otimes \hat{\Pi }_{\rm C} ^{(\textrm{on})} \otimes \hat{\Pi }_{\rm D} ^{(\textrm{on})} 
\right] }{{\cal P}_{\rm det} ^{(2)} } , 
\end{equation}
where
\begin{eqnarray} 
{\cal P}_{\rm det} ^{(2)} &=&1-2e^{-\nu } \frac{1-\lambda ^2 }
{1-\lambda ^2 \{ 1-\eta T_L (1-T)\} } \nonumber \\
&&\hspace{5mm} +e^{-2\nu } \frac{1-\lambda ^2 }{1-\lambda ^2 \{ 1-\eta T_L (1-T)\} ^2 } 
\end{eqnarray}
is the success probability of this event selection. The probability distribution 
of the Bell measurement is 
\begin{eqnarray}
\lefteqn{ {\cal P}_{\rm HD} ^{(2)} (x,p;\lambda ) } \nonumber \\
&=&\langle \Pi (x,p)|\hat{\varrho }_{\rm NG} ^{(2)} |\Pi (x,p)\rangle \nonumber \\
&=&{\cal P}_{11} ^{(2)} (x,p;\lambda ) 
-e^{-\nu } {\cal P}_{10} ^{(2)} (x,p;\lambda ) \nonumber \\
&&\hspace{5mm} -e^{-\nu } {\cal P}_{01} ^{(2)} (x,p;\lambda ) 
+e^{-2\nu } {\cal P}_{00} ^{(2)} (x,p;\lambda ), \label{r-2mode-HD-1}
\end{eqnarray}
where 
\begin{widetext}
\begin{eqnarray}
{\cal P}_{ij} ^{(2)} (x,p;\lambda )
&=&\frac{1}{2\pi {\cal P} _{\rm det} ^{(2)} } 
\frac{1-\lambda ^2 }{(1-\lambda T_L T)^2 
-\lambda ^2 (R_L +\gamma ^{\prime (2)} _i )(R_L +\gamma ^{\prime (2)} _j ) } 
\nonumber \\
&&\hspace{20mm} 
\times \exp \left[ -\frac{1-\lambda ^2 (T_L T+R_L +\gamma ^{\prime (2)} _i )
(T_L T+R_L +\gamma ^{\prime (2)} _j )}{2\{ (1-\lambda T_L T)^2 
-\lambda ^2 (R_L +\gamma ^{\prime (2)} _i )(R_L +\gamma ^{\prime (2)} _j ) \} } 
(x^2 +p^2 )\right] , \label{r-2mode-HD-2}
\end{eqnarray}
and $\gamma ^{\prime (2)} _1 =T_L R$, $\gamma ^{\prime (2)} _0 =(1-\eta )T_L R$. 
With Eqs. (\ref{r-2mode-HD-1}) and (\ref{r-2mode-HD-2}), 
we can calculate the variance of the output state at each port,  
\begin{equation}
{\cal V} ^{(2)} (\lambda )
={\cal V}_{11} ^{(2)} (\lambda )
-e^{-\nu } {\cal V}_{10} ^{(2)} (\lambda ) 
-e^{-\nu } {\cal V}_{01} ^{(2)} (\lambda ) 
+e^{-2\nu } {\cal V}_{00} ^{(2)} (\lambda ), 
\end{equation}
where 
\begin{equation}
{\cal V}_{ij} ^{(2)} (\lambda )=\frac{1}{{\cal P}_{\rm det} ^{(2)} } 
\frac{(1-\lambda ^2 )\{(1-\lambda T_L T)^2 
-\lambda ^2 (R_L +\gamma ^{\prime (2)} _i )(R_L +\gamma ^{\prime (2)} _j )\} }
{\{ 1-\lambda ^2 (T_L T+R_L +\gamma ^{\prime (2)} _i )
(T_L T+R_L +\gamma ^{\prime (2)} _j )\} ^2 } . 
\end{equation}
\end{widetext}
Fig. \ref{r-2mode-HD} shows the probability distribution 
of the output non-Gaussian state ${\cal P}_{\rm HD} ^{(2)} (x,p;\lambda )$ 
along $x$ axis for $\lambda =0.4$ (solid line), 
after the integrating out the variable $p$. We can see that the peak 
is sharper than the input squeezed vacuum state of $\lambda =0.4$ (dotted line). 
The variance of the output non-Gaussian state ${\cal V}(\lambda )$ 
is shown in Fig. \ref{r-2mode-V}. Its variance becomes lower in the range of 
$\lambda \lesssim 0.63$ than that of the input squeezed vacuum state 
(dotted line), which corresponds to the range up to 3.8 [dB] 
in the relative decibel scale. 

\begin{figure}
\centering 
\includegraphics[bb=60 55 555 745, angle=-90, width=.9\linewidth]{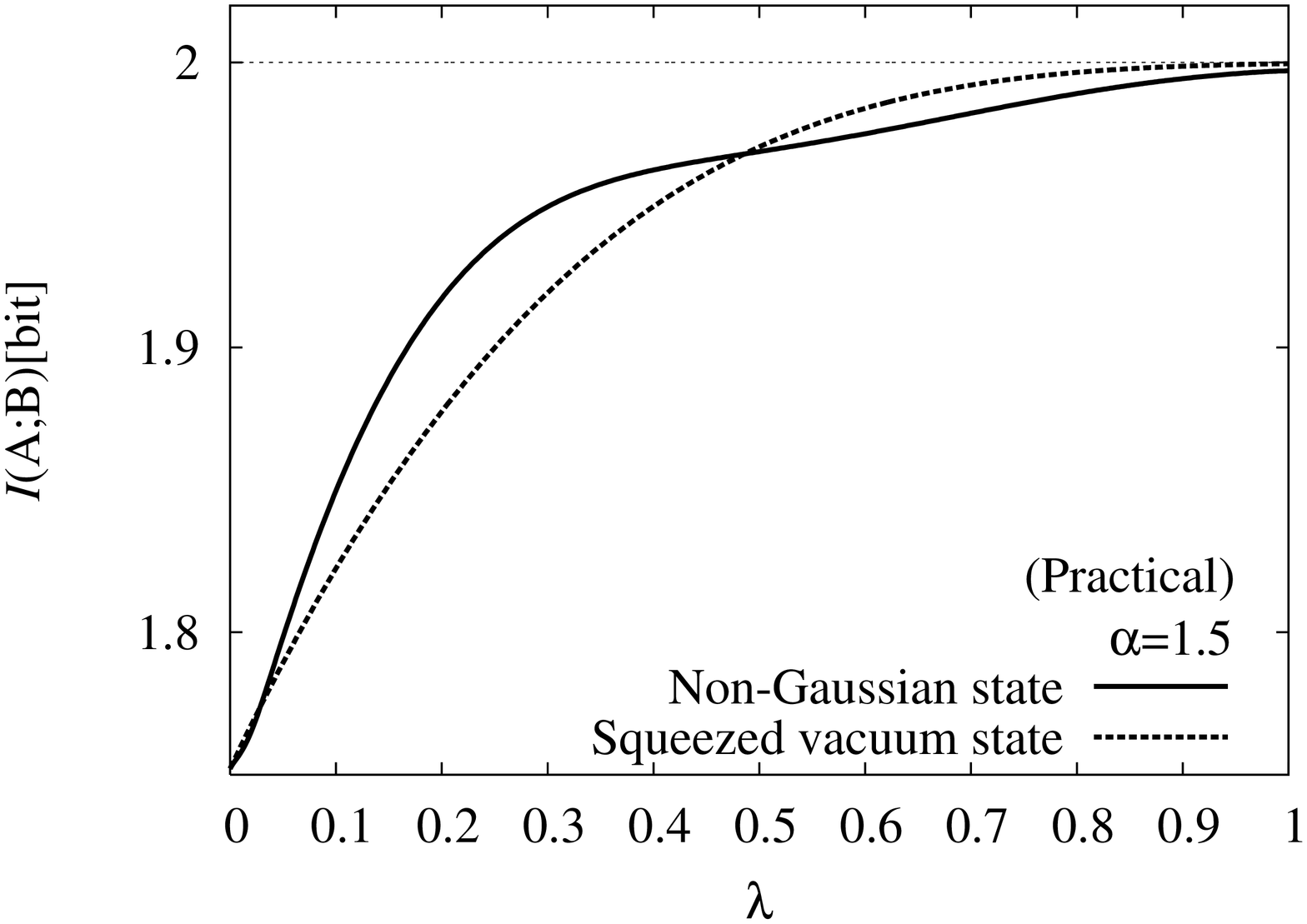}
\caption{\label{r-mutual_info_15} Mutual information via the dense coding channel 
($\alpha =1.5$) of non-Gaussian state (solid line, $T=0.9$) and squeezed vacuum state 
(dotted line) with practical setup ($T_L =0.75$, $\eta =0.6$, $\nu =10^{-3} $). }
\end{figure}

\begin{figure}
\centering 
\includegraphics[bb=60 55 555 745, angle=-90, width=.9\linewidth]{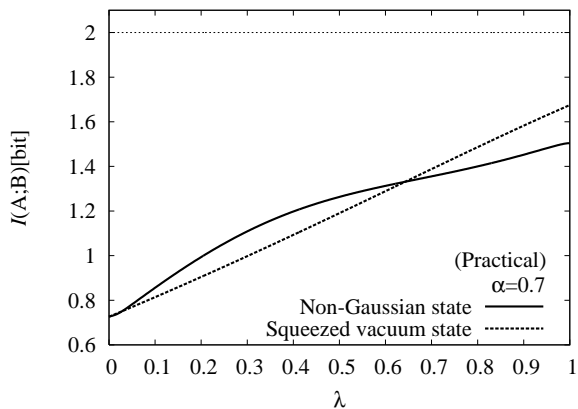}
\caption{\label{r-mutual_info_07} Mutual information via the dense coding channel 
($\alpha =0.7$) of non-Gaussian state (solid line, $T=0.9$) and squeezed vacuum state 
(dotted line) with practical setup ($T_L =0.75$, $\eta =0.6$, $\nu =10^{-3} $). }
\end{figure}

Now, let us consider the dense coding scheme 
with the two-mode non-Gaussian state (\ref{r-2mode-NG}) 
in accordance with Sec. \ref{coding}. 
We assume that the decoherence of the encoded signals in the way 
from Alice to Bob is negligible. Actually we are interested 
in quantifying the effect of non-Gaussian state 
in terms of the mutual information 
rather than in applying the scheme to practical communications. 
The procedure goes in a similar way to the ideal case 
(Fig. \ref{CV_coding}). 
\begin{eqnarray}
\lefteqn{\langle \Pi (x,p)|\hat{U} _{\rm A} (x_s ,p_s )
\hat{\varrho } _{\rm NG} ^{(2)} \hat{U} _{\rm A} ^\dagger (x_s ,p_s ) 
|\Pi (x,p)\rangle } \nonumber \\
&&\hspace{30mm} ={\cal P}_{\rm HD} ^{(2)} (x-x_s ,p-p_s ;\lambda ). 
\label{r-ch-mtrx} 
\end{eqnarray}
With Eq. (\ref{r-ch-mtrx}), we can obtain the channel matrix 
similar to Eq. (\ref{ch-mtrx-component}) and the mutual information can 
be obtained, 
\begin{eqnarray}
\lefteqn{{\cal I} (\textrm{A};\textrm{B}) } \nonumber \\
&&=\sum _{k,l,m,n} P(a_{kl} ){\cal P}(b_{mn} |a_{kl} ) \nonumber \\
&&\hspace{10mm} \times \log _2 \left[ \frac{{\cal P}(b_{mn} |a_{kl} )}
{\displaystyle \sum _{k,l} P(a_{kl} ){\cal P}(b_{mn} |a_{kl} )} \right] , 
\end{eqnarray}
where 
\begin{widetext}
\begin{eqnarray}
{\cal P} (b_{mn} |a_{kl} )&=&\sum _{i,j=0} ^1 \frac{1}{4{\cal P}_{\rm det} ^{(2)} } 
\frac{1-\lambda ^2 }{1-\lambda ^2 (T_L T+R_L +\gamma ^{\prime (2)} _i )
(T_L T+R_L +\gamma ^{\prime (2)} _j )} \nonumber \\
&&\hspace{30mm} 
\times \left( 1+(-1)^{m-k} \textrm{erf}\left[ \Omega ' _{ij} (\lambda ) \alpha \right] 
\right) \left( 1+(-1)^{n-l} \textrm{erf}\left[ \Omega ' _{ij} (\lambda ) \alpha \right] 
\right) , 
\end{eqnarray}
\end{widetext}
and 
\begin{eqnarray}
\lefteqn{\Omega ' _{ij} (\lambda )} \nonumber \\
&&=\sqrt{
\frac{1-\lambda ^2 (T_L T+R_L +\gamma ^{\prime (2)} _i )
(T_L T+R_L +\gamma ^{\prime (2)} _j )}
{(1-\lambda T_L T)^2 -\lambda ^2 (R_L +\gamma ^{\prime (2)} _i )
(R_L +\gamma ^{\prime (2)} _j )} } . \nonumber \\
\end{eqnarray}
In Figs. \ref{r-mutual_info_15} and \ref{r-mutual_info_07}, 
the mutual information of $\alpha =1.5$ and $\alpha =0.7$ 
are indicated, respectively. 
The mutual information of the non-Gaussian state increases 
over that of the original squeezed vacuum state, 
in the range of $\lambda \lesssim 0.47$ 
for $\alpha =1.5$ and in $\lambda \lesssim 0.65$ for $\alpha =0.7$. 
This gain may be attributed to the effective increase of the entanglement, 
and will be seen even under practical situation.

\section{Discussion and conclusion} \label{discussion}

In this paper, we have studied the non-Gaussian operations 
induced by the measurement with the on-off detectors  
on the single- and two-mode \textit{Gaussian} squeezed vacuum states. 
Our scheme is the Mach-Zehnder interferometer, 
where the non-Gaussian state measured at the two output ports 
with respect to the single quadrature in each port. 
This setup was originally motivated by the naive intuition 
that the useful non-Gaussian states induced by the measurement 
should exhibit smaller variances in appropriate quadratures than 
the ones of the originals, irrespective of the mixedness of the 
state and the amount of the deviation from the Gaussian state.  
In both the single- and two-mode cases, 
the homodyne probability distributions at the two output ports 
show the single main peak which is very close to the Gaussian.  
The effect of the non-Gaussian operations based on the on-off detector 
and the usefulness of the resulting states may be characterized 
simply by the reduction of the variances. 
In the two-mode case, the two variances measured 
at the two output ports can still be regarded 
as a measure of the quantum correlation 
in the non-Gaussian bipartite system induced 
by the on-off detector. This sort of evolution 
is of course not rigorous mathematically. 
More satisfactory theories need to be developed.

Our analysis includes possible practical imperfections, 
such as the linear loss in optical paths, 
finite quantum efficiency, and nonzero dark counts 
of the on-off detector. We have seen that 
although they degrade the output non-Gaussian state, 
the gains in terms of the variances and the mutual information 
of the dense coding scheme still remains. One of the important effects 
that has not been involved in this paper is the mode mismatch 
between the field measured by the homodyne detector 
and the field of trigger photons measured by the on-off detector. 
Ideally, the field mode of trigger photons must be in a mode 
which overlaps perfectly the local oscillator mode 
in the homodyne detector, with respect to the spatial, 
temporal, and frequency domains. Otherwise one cannot 
select the matched mode, and is led to the degradation of the gains. 
This aspect was studied in \cite{Grosshans01}. An alternative analysis 
based on the present model will be presented elsewhere.

Then, as an operational measure of non-Gaussian operation, 
we have studied the mutual information in quantum dense coding scheme. 
The dense coding is one of the entanglement-assisted schemes, 
and the improvement of the mutual information is considered as the gain 
assisted by the enhanced entanglement via non-Gaussian operation. 
The mutual information in the dense coding scheme has 
some advantageous features. Firstly, it is a well-established 
measure in information theory, which is obtained by specifying 
the channel matrix, and has a clear operational meaning 
in a multiple use of  the channel. 
Secondly, it can be a rigorous measure for the obtained gain. 
Actually, the gain in the mutual information usually gets lost 
even by a small amount of imperfections. So the observed gain will 
ensure that the system really works. Finally, to evaluate 
the mutual information is also suitable for practical use. In fact, 
the corresponding experimental setup is relatively easy 
in laboratory compared with, for example, quantum teleportation. 
This measure would be the first quantity to evaluate 
when the non-Gaussian state is experimentally available.

On the other hand, more strict theories for measure of \textit{mixed} 
entangled state has to be developed, which is earnestly desired 
both theoretically and experimentally.

\begin{acknowledgments}
The authors would like to M.~Ban and S.~L.~Braunstein 
for valuable discussions. 
\end{acknowledgments}

\end{document}